# Linear Cellular Automata as Discrete Models for Generating Cryptographic Sequences

Pino Caballero-Gil[1], Amparo Fúster-Sabater[2], Oscar Delgado-Mohatar[2]

[1]Faculty of Mathematics. University of La Laguna
38271. La Laguna, Tenerife, Spain.
pcaballe@ull.es
[2]Institute of Applied Physics. Spanish High Council for Scientific Research.
Serrano, 144, 28006, Madrid, Spain.
{amparo, oscar.delgado}@iec.csic.es

**Abstract.** In this paper, we develop a new cellular automata-based linear model for several nonlinear pseudorandom number generators with practical applications in symmetric cryptography. Such a model generates all the solutions of linear binary difference equations as well as many of these solutions are pseudo-random keystream sequences. In this way, a linear structure based on cellular automata may be used to generate not only difference equation solutions but also cryptographic sequences. The proposed model is very simple since it is based exclusively on successive concatenations of a basic linear automaton.

## 1    Introduction

Cellular Automata (CA) are discrete dynamical systems that operate on a uniform, regular lattice, and are based on simple local interactions among its components. Consequently, their main properties are discreteness (in space, time and values), local interaction, homogeneity, and parallel evolution (Wolfram, 1986). A large number of research papers on CA are published every year. The main reason behind the popularity of CA is the enormous potential they hold in modelling complex systems, in spite of their simplicity. These uniform arrays of identical cells in an *n*-dimensional space may be characterized by four different parameters: cellular geometry, neighbourhood specifications, number of states per cell, and transition rules. In this work, our interest is concentrated on one-dimensional binary CA with three site neighborhood and *linear* transition rules. In addition, CA here considered will be *hybrid* (different cells evolve under different transition rules) and *null* (cells with null content are adjacent to the automaton extreme cells).

The second goal of this paper is stream cipher, which is the fastest encryption procedure nowadays. Consequently, stream cipher procedures are implemented in many practical applications (e.g., the algorithms A5 in GSM communications GSM), and the encryption system E0 in Bluetooth specifications (Bluetooth)). From a short secret key (known only by the two interested parties) and a public algorithm (the sequence generator), stream cipher procedures consist in generating long sequences of seemingly random bits, which are pseudo-random sequences. In cryptographic terms, such sequences are called keystream sequences.

In the literature we can find many different families of pseudo-random sequences based on Linear Feedback Shift Registers (LFSRs). The output sequences of such linear registers are combined by means of nonlinear functions in order to produce keystream sequences of cryptographic application. They can be generated in two different ways:

1. By a LFSR controlled by another LFSR, which may be the same one (e.g., multiplexed sequences (Jennings, 1983), clock-controlled sequences (Beth and Piper, 1985), cascaded sequences (Gollmann and Chambers, 1989), and shrinking generator sequences (Coppersmith et al, 1994)).
2. By one or more than one LFSR and a feed-forward nonlinear function (e.g., Gold-sequence family, Kasami sequence families, GMW sequences, Klapper sequences and No sequences). See (Gong, 1995) and the references cited therein.

In the present work, it is shown that one-dimensional linear CA based on rules 90/150 generate all the solutions of linear difference equations with binary constant coefficients. Some of these solutions correspond to the sequences produced by the aforementioned keystream generators. In this way, we obtain simple CA that not only generate all the solutions of a kind of difference equations but also are linear models of nonlinear cryptographic sequence generators. Due to the linearity of the CA transition rules, modelling these CA-based structures is simple and efficient. To our knowledge, there are no CA-based linear models able to produce well-known keystream sequences currently obtained from LFSR-based generators.

This work is organized as follows. In the next section, fundamentals and basic notation of linear binary difference equations and one-dimensional linear hybrid CA are introduced. A generalization of such difference equations is provided in section 3. Many solutions of these equations are cryptographic sequences generated by the corresponding CA as it is shown in section 4. Finally, an illustrative example and conclusions complete the paper.

## 2  Background

In this section, the two basic structures considered within this paper (linear difference equations and one-dimensional linear hybrid CA) are briefly introduced.

## 2.1 Linear Binary Difference Equations

Throughout this work, the following kind of linear difference equations with binary coefficients will be considered:

$$(E^r \oplus \sum_{j=1}^{r} c_j E^{r-j}) a_n = 0, \quad n \geq 0$$

where $a_n \in GF(2)$ is the *n*-th term of a binary sequence $\{a_n\}$ that satisfies the previous equation. $E$ is the shifting operator that operates on the terms $a_n$ of a solution sequence (i.e. $E^j a_n = a_{n+j}$). The coefficients $c_j$ are constant binary coefficients $c_j \in GF(2)$, $r$ is an integer and the symbol $\oplus$ represents the XOR logic operation. The *r*-degree characteristic polynomial of the equation (1) is:

$$P(x) = x^r + \sum_{j=1}^{r} c_j x^{r-j} \qquad (2)$$

and specifies the linear recurrence relationship of the sequence $\{a_n\}$. This means that its *n*-th term, $a_n$, can be written as a linear combination of the previous terms:

$$a_n \oplus \sum_{j=1}^{r} c_j a_{n-j} = 0, \quad n \geq r. \qquad (3)$$

In the sequel, $P(x)$ will be a primitive binary coefficient polynomial (Key, 1976) so that if $\alpha$ is one of its roots, then

$$\alpha, \alpha^2, \alpha^{2^2}, \ldots, \alpha^{2^{(r-1)}} \in GF(2^r) \qquad (4)$$

are the *r* different roots of such a polynomial. For more details see (Lidl and Niederreiter, 1986). In this case, it can be proved (Key, 1976) that the solutions of equation (1) are sequences of the form:

$$a_n = \sum_{j=0}^{r-1} A^{2^j} \alpha^{2^j n}, \quad n \geq 0 \qquad (5)$$

where $A$ is an arbitrary element in $GF(2^r)$. That is to say, $\{a_n\}$ is a Pseudo-Noise sequence (*PN*-sequence) of characteristic polynomial $P(x)$ and period $2^r - 1$ whose starting point is determined by the value of $A$. If $A = 0$, then the solution of equation (1) is the identically null sequence.

## 2.2 One-Dimensional Linear Hybrid CA

Now, our attention is focused on one-dimensional binary linear hybrid CA with three site neighborhood. In fact, there are eight of such transition rules among which only two (rule 90 and rule 150) lead to non trivial structures. Both rules are defined as follows (Kari, 2005):

$$\text{Rule 90} \qquad\qquad \text{Rule 150}$$
$$b_{n+1}^{k} = b_{n}^{k-1} \oplus b_{n}^{k+1} \qquad\qquad b_{n+1}^{k} = b_{n}^{k-1} \oplus b_{n}^{k} \oplus b_{n}^{k+1}$$

Indeed, at time $n + 1$ the content of the $k$-th cell, $b_{n+1}^{k} \in GF(2)$, depends on the content at time $n$ of either two different cells (rule 90) or three different cells (rule 150), with $k = 1, ..., L$, where $L$ is the length of the automaton. Moreover, the state of the automaton at time $n$ is the binary content of the $L$ cells at such an instant. A natural form of representation for this kind of automata is given by an $L$-tuple $\Delta_L = (d_1, d_2, ..., d_L)$ where $d_k = 0$ if the $k$-th cell verifies rule 90 while $d_k = 1$ if the $k$-th cell verifies rule 150. In addition, $\Delta_k = (d_1, d_2, ..., d_k)$ for $k = 1, ..., L$ denotes the corresponding sub-automaton of length $k$.

Given a primitive polynomial $Q(x)$, the Cattell and Muzio synthesis algorithm (Cattell et al, 1999) provides us with a pair of reversal linear 90/150 CA whose characteristic polynomial is $Q(x)$. Therefore, a one-dimensional binary linear 90/150 cellular automaton of primitive characteristic polynomial $P(x)$ given by (2) will generate the *PN*-sequence defined by equation (5), see (Fúster-Sabater and Caballero-Gil, 2007). As an example, Table 1 depicts in bold the *PN*-sequence obtained either:

1. As a solution of equation (1) with characteristic polynomial $P(x) = x^3 + x^2 + 1$, $r = 3$ and $A = 1$ ($A \in GF(2^3)$)
$$a_n = 1\alpha^n \oplus 1\alpha^{2n} \oplus 1\alpha^{4n}, \quad n \geq 0 \qquad (6)$$

2. As the sequence generated by the pair of reverse CA (e.g., (150 90 90) and (90 90 150)) starting at the initial states (1,0,1) and (1,1,0), respectively. At the remaining CA cells, shifted versions of the same *PN*-sequence are generated.

**Table 1.** The same *PN*-sequence obtained either as a difference equation solution or as a sequence generated by two reversal linear CA

| Differen. eq. sol. | CA: 150 | 90 | 90 | 150 | 90 | 90 |
|---|---|---|---|---|---|---|
| **1** | 1 | 0 | 1 | 1 | **1** | 0 |
| **1** | 1 | 0 | 0 | 1 | **1** | 1 |
| **1** | 1 | 1 | 0 | 1 | **0** | 0 |
| **0** | 0 | 1 | 1 | 0 | **1** | 0 |
| **1** | 1 | 1 | 1 | 1 | **0** | 1 |
| **0** | 0 | 0 | 1 | 0 | **0** | 1 |
| **0** | 0 | 1 | 0 | 0 | **1** | 1 |

## 3  Generalization

Let us generalize the difference equations in subsection 2.1 to a more complex kind of linear difference equations whose roots have a multiplicity greater than 1. In fact, we are going to consider equations of the form:

$$(E^r + \sum_{j=1}^{r} c_j E^{r-j})^p a_n = 0, \quad n \geq 0 \tag{7}$$

$p$ being an integer $> 1$. The characteristic polynomial $P_M(x)$ of this kind of equations is:

$$P_M(x) = P(x)^p = (x^r + \sum_{j=1}^{r} c_j x^{r-j})^p \tag{8}$$

In this case, the roots of $P_M(x)$ are the same as those of the polynomial $P(x)$, that is, $(\alpha, \alpha^2, \alpha^{2^2}, \ldots, \alpha^{2^{(r-1)}})$, but with multiplicity $p$. Now the solutions of (7) are:

$$a_n = \sum_{i=0}^{p-1} \left( \binom{n}{i} \sum_{j=0}^{r-1} A_i^{2^j} \alpha^{2^j n} \right) \tag{9}$$

where $A_i \in GF(2^r)$ and the $\binom{n}{i}$ are binomial coefficients modulo 2. According to (5), the term $\sum_{j=0}^{r-1} A_i^{2^j} \alpha^{2^j n}$ represents the *n-th* element of the same *PN*-sequence as before whose starting point is determined by $A_i$. Thus, $\{a_n\}$ is just the bitwise sum of $p$ times the same *PN*-sequence starting at different points and weighted by different binomial coefficients. In fact, each binomial coefficient defines a succession of binary values with a constant period $T_i$. Table 2 shows the sequences and values of $T_i$ for the first coefficients $\binom{n}{i}$ ($i = 0, \ldots, 7$).

**Table 2.** Binomial coefficients, binary values and periods $T_i$

| Binomial coeff. | Binary values | $T_i$ |
|---|---|---|
| $\binom{n}{0}$ | 1,1,1,1,1,1,1,1,1, . . . | $T_0 = 1$ |
| $\binom{n}{1}$ | 0,1,0,1,0,1,0,1,0,1, . . . | $T_1 = 2$ |
| $\binom{n}{2}$ | 0,0,1,1,0,0,1,1,0,0, . . . | $T_2 = 4$ |
| $\binom{n}{3}$ | 0,0,0,1,0,0,0,1,0,0, . . . | $T_3 = 4$ |
| $\binom{n}{4}$ | 0,0,0,0,1,1,1,1,0,0, . . . | $T_4 = 8$ |
| $\binom{n}{5}$ | 0,0,0,0,0,1,0,1,0,0, . . . | $T_5 = 8$ |
| $\binom{n}{6}$ | 0,0,0,0,0,0,1,1,0,0, . . . | $T_6 = 8$ |
| $\binom{n}{7}$ | 0,0,0,0,0,0,0,1,0,0, . . . | $T_7 = 8$ |

Remark that the choice of the coefficients $A_i$ determines the characteristics of the sequences $\{a_n\}$ that are solutions of equation (7). Indeed, the period $T$ of $\{a_n\}$ depends on the periods $T_i \cdot (2^r - 1)$ of the $p$ sequences that are summed in (9). The linear complexity $LC$ of $\{a_n\}$ is related with the number of roots with their corresponding multiplicities weighted by $A_i$ that appear in (9). The number $N$ of different sequences {$a_n$} is related with the number of different $p$-tuples of values of $A_i$.

This kind of difference equations given by (7) is crucial because many binary sequences of cryptographic application (those ones referred in the introduction) have characteristic polynomials given by equation (8). That is, for example, the case of cryptographic interleaved sequences (Gong, 1995). Consequently, many cryptographic sequences are solutions of linear difference equations. In this way, it would be very convenient to have a simple CA-based linear model able to compute all the solutions of these difference equations, among them we could find a great variety of cryptographic sequences. Next section tackles this problem.

## 4    Linear Difference Equation Solutions by means of CA

Since the characteristic polynomial of the considered equations is $P_M(x) = P(x)^p$, it seems quite natural to construct the solutions of such equations by concatenating $p$ times the basic automaton of characteristic polynomial $P(x)$. The following result is a concrete formalization of this idea.

**Theorem 1.** *Let C be a linear 90/150 cellular automaton of length L, binary codification ($d_1, d_2, ..., d_{L-1}, d_L$) and characteristic polynomial $P(x)$. Let $\tilde{C}$ be the reversal version of C, with binary codification ($d_L, d_{L-1}, ..., d_2, d_1$), and the same length and polynomial as those of C. Then, the 2L-tuple defined by ($d_1, d_2, ..., \overline{d_L}, \overline{d_L}, ..., d_2, d_1$), represents the linear 90/150 cellular automaton of length 2L and characteristic polynomial $P(x)^2$.*

The proof of this theorem is based on the recurrence relationship for the characteristic polynomials of the successive sub-automata of a given automaton (Cattell et al, 1999).

The result can be iterated a number of times for successive polynomials and rule vectors:

$$
\begin{aligned}
P(x) &\leftrightarrow \Delta_L = (d_1, d_2, ..., d_L) \\
P(x)^2 &\leftrightarrow \Delta_{2L} = (d_1, d_2, ..., \overline{d_L}, \overline{d_L}, ..., d_2, d_1) \\
P(x)^{2^2} &\leftrightarrow \Delta_{2^2 L} = (d_1, d_2, ..., \overline{d_L}, \overline{d_L}, ..., d_2, \overline{d_1}, \overline{d_1}, d_2, ..., \overline{d_L}, \overline{d_L}, ..., d_2, d_1) \\
&\vdots \leftrightarrow \vdots
\end{aligned}
$$

Note that the basic automaton is concatenated with its reversal version after the complementation of the last rule. Successive applications of this result provide us with CA whose characteristic polynomials are: $P(x)^2, P(x)^{2^2}, P(x)^{2^3}, ..., P(x)^{2^q}$ of lengths $2L, 2^2L, 2^3L, ..., 2^qL$, respectively. Remark that for every $P(x)$ there are two reverse basic automata that may be used in the concatenation procedure. Consequently, for $2^{q-1} < p \leq 2^q$ the two automata built as in Theorem 1 will produce for different CA initial states all the sequences {$a_n$} with characteristic polynomial $P(x)^p$ that satisfy the difference equation.

## 5 Illustrative Example

Let us see a simple example illustrating the previous sections. Consider a pair of reverse CA $\Delta_5$= (1,0,0,0,0) and $\Delta^*_5$= (0,0,0,0,1) of length $L$=5 associated to the characteristic polynomial $P(x) = x^5+x^4+x^2+x+1$. If $P_M(x)= P(x)^p$ with $p$= 4, then one of the CA obtained by concatenation will be: $\Delta_{20}$= (1, 0, 0, 0, 1, 1, 0, 0, 0, 0, 0, 0, 0, 0, 1, 1, 0, 0, 0, 1). Now different choices of $A_i$ (not all zero) will allow us to generate all the solutions of the difference equation (7) with the previous parameters.

1. If $A_0 \neq 0$ and $A_i$= 0 $\forall i > 0$, then the cellular automaton will produce $N_0$= 1 sequence, which is a unique *PN*-sequence of period $T_0$= 31, linear complexity $LC_0$= 5 and characteristic polynomial $P(x)$. In addition, the automaton cycles through doubly symmetric states of the form:

    $(a_0, a_1, a_2, a_3, a_4, a_4, a_3, a_2, a_1, a_0, a_0, a_1, a_2, a_3, a_4, a_4, a_3, a_2, a_1, a_0)$

    with $a_i \in GF(2)$. The 31 doubly symmetric states are concentrated into the same cycle.

2. If $A_1 \neq 0$ and $A_i$= 0 $\forall i > 1$, then the cellular automaton will produce $N_1$= 16 different sequences of period $T_1$= 62, linear complexity $LC_1$= 10 and characteristic polynomial $P(x)^2$. Moreover, the automaton cycles through symmetric states of the form:

    $(a_0, a_1, a_2, a_3, a_4, a_5, a_6, a_7, a_8, a_9, a_9, a_8, a_7, a_6, a_5, a_4, a_3, a_2, a_1, a_0)$

    with $a_i \in GF(2)$. In fact, there are $2^{10}$-32= 992 symmetric states distributed in 16 cycles of 62 states each of them.

3. If $A_2 \neq 0$ and $A_i$= 0 $\forall i > 2$, then the cellular automaton will produce $N_2$= 256 different sequences of period $T_2$= 124, linear complexity $LC_2$= 15 and characteristic polynomial $P(x)^3$. Moreover, the automaton cycles through several repetitive states of the form:

    $(a_0, a_1, a_2, a_3, a_4, a_5, a_6, a_7, a_8, a_9, a_0, a_1, a_2, a_3, a_4, a_5, a_6, a_7, a_8, a_9)$

    with $a_i \in GF(2)$.

4. If $A_3 \neq 0$, then the cellular automaton will produce $N_3$= 8192 different sequences of period $T_3$= 124, linear complexity $LC_3$= 20 and characteristic polynomial $P(x)^4$. In addition, the automaton cycles through the states not included in the previous cycles.

In brief, a simple linear structure based on CA allows us by successive concatenations to compute in a natural way all the solutions of linear binary difference equations, some of which are sequences of cryptographic application.

## 6 Conclusion

This paper has shown that all the solutions of linear binary difference equations can be realized by means of linear models based on 90/150 cellular automata. It is remarkable that some of these solutions have a straight cryptographic application in stream ciphers because they are keystream sequences produced with pseudo-random generators. In this way, very popular cryptographic sequence generators conceived and designed as nonlinear generators are here linearized in terms of cellular automata. To our knowledge, these are the first CA-based linear models generating well-known keystream sequences. The linearization procedure is simple and can be applied to cryptographic examples in a range of practical application since the hardware implementation of the developed CA-based models is easy and very adequate for FPGA logic. This characteristic makes it suitable for developments where time execution is relevant as in communication systems with high transmission rates.

## Acknowledgements

This research has been supported by the Spanish Ministry of Science and Innovation under Project TIN2008-02236/TSI, and developed for the project HESPERIA (www.proyecto-hesperia.org) under programme CENIT supported by Centro para el Desarrollo Tecnológico Industrial (CDTI) and the companies: Soluziona, Unión Fenosa, Tecnobit, Visual-Tools, BrainStorm, SAC and TechnoSafe.